\documentclass[lettersize,journal]{IEEEtran}
\usepackage{amsmath,amsfonts}
\usepackage{algorithmic}
\usepackage{algorithm}
\usepackage{array}
\usepackage[caption=false,font=normalsize,labelfont=sf,textfont=sf]{subfig}
\usepackage{textcomp}
\usepackage{stfloats}
\usepackage{url}
\usepackage{verbatim}
\usepackage{graphicx}
\usepackage{cite}
\begin{document}
\title{A Method for Constructing Wavelet Functions on the Real Number Field}

\author{Ning Li and Lezhi Li}

\maketitle

\begin{abstract}
A general method  to construct wavelet function on real number field is proposed in this article,which is based on finite length sequence.This finite length sequence is called the seed sequence, and the corresponding wavelet function is called the seed sequence wavelet function.The seed sequence wavelet function is continuous and energy concentrated in both time and frequency domains.That is, it has a finite support set in both time and frequency domains. It is proved that if and only if the seed sequence has 0 mean value, the interpolation function satisfies the admissible condition of wavelet function.The conditions corresponding to the higher order vanishing moment of the seed sequence wavelet function are also given in this article. On this basis, the concept of random wavelet function is proposed, and the condition of the regularity of random wavelet is discussed. 
\end{abstract}


\section{Introduction}
\IEEEPARstart{T}{ime}-frequency domain analysis (TF) plays a very important role in processing non-stationary time series signals, and wavelet analysis is an important time-frequency domain analysis method. \cite {huang1998empirical}\cite {Wavelet_Analysis_CHEN}
Since J.Morlet proposed the concept of wavelet analysis, scientists have found many different functions as wavelet generating functions\cite{Morletsound1987}\cite{Improved_Morlet}\cite{TenLectures}. Generally, the wavelet generating function is called the wavelet function.

Time-frequency domain analysis method is a method to map one-dimensional time signal to two-dimensional time-frequency plane
\cite{trackingbasedonMorlet}.The essence of wavelet transform is "raising dimension" from the original one-dimensional time domain to the two-dimensional "scale-displacement" plane to study signals. "Increasing dimension" will certainly bring a certain "redundancy"\cite{daubechiesOrthonormalbasis}\cite{Time-Frequency/Time-ScaleAnalysis1999}.
In other words, it is only necessary to use certain sets of points on the two-dimensional "scale-displacement" plane to establish a correspondence with the one-dimensional time domain signal.

Based on this understanding, discrete wavelet is proposed, and the corresponding relationship between the one-dimensional continuous time signal and the set of discrete points in the two-dimensional scale-displacement plane is established
\cite{Mallatmultiresolution}.Discrete wavelets can also be regarded as a set of functions derived from wavelet generating functions, and such a set of functions is used as a wavelet base\cite{OrthogonalitycriteriaLian}\cite{daubechiesOrthonormalbasis}.
Furthermore, the framework of multi-resolution analysis is established. In the framework of multi-resolution analysis, the direct sum of the wavelet space and the corresponding scale space is another scale space\cite{MULTIRESOLUTION_APPROXIMATION}.
The signal can be decomposed into detail part W and large scale approximation part V. Then, the large scale approximation can be further decomposed. By repeating this, you can get detailed parts and approximation parts at any scale (or resolution).
Therefore, the wavelet function can be obtained by evaluating the difference between two different scale functions. The wavelet reflects the small change between adjacent scales, that is, the detailed part of the signal
\cite{TenLectures}\cite{Mallatmultiresolution}.
A scale function passes through the corresponding low-pass filter to get another adjacent scale function, and the scale function passes through the high-pass filter to get the corresponding wavelet function. The low-pass filter and high-pass filter form a pair.This is the principle of the two-scale difference equation
\cite{Multiresolution_Signal2001}\cite{ref19}.
After establishing the correspondence between the coefficients of the high-pass filter and the coefficients of the low-pass filter, the correspondence between the scale function and the wavelet function can be determined.
B-spline curve has the property of maintaining geometric shape to a certain extent, so B-spline curve is suitable for some kind of scale function.  Corresponding B-spline wavelet function can be generated based on B-spline scale function.
\cite{Third-order B-spline wavelet}\cite{Multiresolution_Signal2001}.

In recent years, a new signal analysis method in time-frequency domain, the mode decomposition method has been developed. It mainly includes EMD, VMD and OMD
\cite{huang1998empirical}
\cite{dragomiretskiy2013variational}
\cite{LiOrthogonalModeDecomposition}.
The modes mentioned here are essentially narrow-band functions. According to the definition of mode, it can be divided into two categories. One is the narrow-band function with periodic property, and the other is the narrow-band function without periodic property.
Because narrowband function without periodic properties have local support sets in the time domain, they have the properties of wavelet functions\cite{LiOrthogonalModeDecomposition}.In this sense, the mode decomposition method and the wavelet analysis method are deeply related\cite{LiOrthogonalModeDecomposition}\cite{daubechies2011synchrosqueezed}.
The above actually discusses the existing methods of generating wavelet functions, including classical wavelet functions, as well as generating wavelet functions based on multi-resolution analysis frameworks or mode decomposition method.
The number of classical wavelet functions available is limited. However, the wavelet function generated based on multi-resolution analysis framework generally has no uniform analytic expression, such as B-spline wavelet function.

The wavelet function should satisfy the admissibility condition\cite{admissibleconditionswavelet2005}.
In some applications, the wavelet function is required to have a high order vanishing moment, that is,there are certain requirements on the smoothness of the wavelet function\cite{Regularitycompactlysupported2010}.
However, it is very difficult to find continuously differentiable wavelet functions with compact supports, and even more difficult if orthogonality is also required. In order to satisfy the requirement of orthogonality, multi-wavelet is proposed\cite{ConstructionofBiorthgonal}.

In this paper, a new method is used to construct the wavelet function on the real number field, which is based on the interpolation function of the real sequence of finite length.
The finite length sequence is called the seed sequence and the generated wavelet function is called the seed sequence wavelet function.The seed sequence wavelet function is continuous and differentiable, and it has a finite support set in both the time domain and the frequency domain. We prove that if and only if the seed sequence has zero mean, its interpolation function satisfies the admissible condition of the wavelet function. The conditions corresponding to the higher order vanishing moment of the wavelet function of the seed sequence are also given. Based on this, the concept of random wavelet function is proposed, and the condition of the regularity of random wavelet is discussed.

\section{Constructing the wavelet function}
\label{sec:construct_wavelet }
\textbf{Theorem 1:} $\forall$ sequence of real numbers $\boldsymbol{u}$:
\vspace{-1mm}
$$\boldsymbol{u}=[u(-l),u(-l+1),\hdots,u(0),\hdots,u(l-1),u(l)]$$   
\indent Considering $\boldsymbol{u}$ as a time series, and the sampling interval between the elements in the series is $\Delta$. The total length of time is $T=2l\Delta$, and 
 $\sum_{k=-l}^{l}{u(k)=0}$. 
 
We construct the interpolation function $\Psi_u (t)$  with the time series $\boldsymbol{u}$ according to the formula (\ref{eq:welet_construct1}) :

\begin{equation}
\label{eq:welet_construct1}
    \Psi_u(t) = \sum_{k=-l}^{l}{u(k) \frac{\sin \Omega_{\Delta}(t - k\Delta) }{\Omega_{\Delta}(t - k\Delta)}}
\end{equation}

In the formula(\ref{eq:welet_construct1}), $t \in (-\infty, \infty)$, $\Omega_{\Delta} = \pi/\Delta$. 

Then $\Psi_u(t)$ is a wavelet function on the real field. We call the sequence $\boldsymbol{u}$ the seed sequence, and the wavelet function $\Psi_u (t)$ the seed sequence wavelet function.

\vspace{1mm}
\textbf{Proof:}
\begin{enumerate}
\item
From formula(\ref{eq:welet_construct1}), we know that:

\begin{equation}
  \Psi_{u}(k \Delta) =
    \begin{cases}
      u(k) & \text{if  k >= -l and k<=l }\\
      0 & \text{elif  k <- l or k > l}
    \end{cases}   \nonumber     
\end{equation}

  $\Psi_{u}$ has a limited support interval $[-l\Delta,l\Delta]$. Outside this interval,
$\Psi_{u}$Rapidly oscillates to zero.

\item
The Fourier transform of $\Psi_u(t)$ is :
\begin{align}
F(\omega) & = \int_{-\infty}^{\infty} \Psi_u(t) e^{-j\omega t} \,dt \nonumber\\ 
& = \int_{-\infty}^{\infty} \sum_{k=-l}^{l}{u(k) \frac{\sin \Omega_{\Delta}(t - k\Delta) }{\Omega_{\Delta}(t - k\Delta)}} e^{-j\omega t} \,dt \nonumber \\
& = \sum_{k=-l}^{l}{u(k) \int_{-\infty}^{\infty} \frac{\sin \Omega_{\Delta}t'}{\Omega_{\Delta}t'}} e^{-j\omega (t' + k\Delta)} \,dt' \nonumber \\ 
& = \sum_{k=-l}^{l}{u(k) e^{-j\omega k\Delta}} \int_{-\infty}^{\infty} \frac{\sin \Omega_{\Delta}t}
{\Omega_{\Delta}t} e^{-j\omega t}\,dt \nonumber \\
&= U_u(\omega)U_{\Omega_{\Delta}}(\omega) \label{fourier_psi} 
\end{align} 

\begin{flalign}
& U_{\Omega_{\Delta}}(\omega) = \int_{-\infty}^{\infty} \frac{\sin \Omega_{\Delta}t} {\Omega_{\Delta}t}e^{-j\omega t} \,dt = 
    \begin{cases}
      \Delta & |\omega| <= \Omega_{\Delta} \\
      0      & |\omega| > \Omega_{\Delta} 
    \end{cases} \label{u_delta}
\\
& U_u(\omega) = \sum_{k=-l}^{l}{u(k) e^{-j\omega k\Delta}} \label{u_u}
\end{flalign}

According to the given condition $\sum_{k=-l}^{l}{u(k)=0}$, we know that: 

$$ U_u(0)=0, F(0)=0, \lim_{\omega \to 0} \frac{|F(\omega)|^2}{\omega}=0 $$

So the improper Integral $\int_{0}^{\infty} \frac{|F(\omega)|^2}{\omega} \,dt $ is convergent, that is:

$$0<C_{\Psi}=\int_{0}^{\infty} \frac{|F(\omega)|^2}{\omega}\,dt <\infty $$

The function $\Psi_u(t)$ satisfies the admissible condition.

Q.E.D.

\end{enumerate}
\vspace{1mm}

From the above proof process, it can be seen that $\Psi_u(t)$ has a finite support set in both the time domains and the frequency domains.
\vspace{1mm}

\textbf{Example 1:}
Haar wavelet is a simple and easy to understand wavelet function, which is widely used in signal processing and other fields, but Haar wavelet is discontinuous.

Taking $\frac{1}{2}$ as the sampling period and sampling the Haar wavelet function at sampling point ${\{\frac{1}{4},\frac{3}{4}\}}$ , the sequence ${\{1,-1\}}$ can be obtained.
The sequence ${\{1,-1\}}$ accords with the requirements of the seed sequence proposed in this paper. Therefore, a new continuous wavelet function can be constructed, as below:
$$\Psi_u (t)=\frac{sin(2\pi(t - \frac{1}{4}))}{2\pi(t - \frac{1}{4})}-\frac{sin(2\pi(t - \frac{3}{4}))}{2\pi(t - \frac{3}{4})}\label{psi}$$

\begin{figure}[t]
\centering
\includegraphics[width=2.5in]{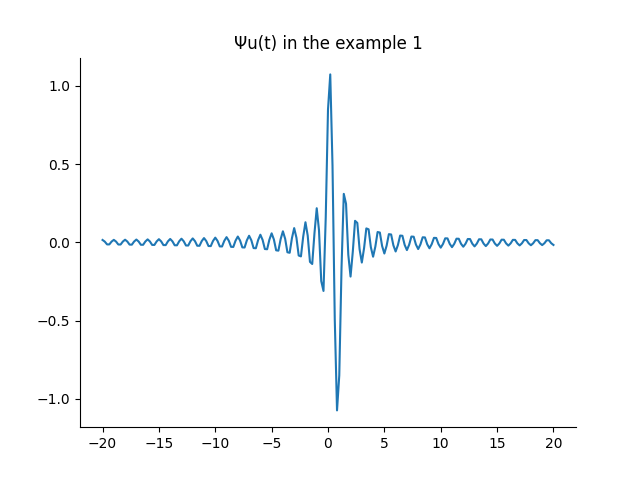}
\caption{The continuous wavelet is obtained by modifying Harr wavelet}
\label{fig_1}
\end{figure}

The wavelet function $\Psi_u (t)$ in example 1 is shown in Fig.\ref{fig_1}.

In Fig.\ref{fig_1}, the primary value interval of $\Psi_u (t)$ is [0,1].The positive peak is 1 and the negative peak is -1. Example 1 shows the wavelet function generated by the seed sequence $\{1,-1\}$.

\section{Vanishing moment of  seed sequence wavelet}
\label{sec:Vanishing_Moments }

If the wavelet function $\Psi_u (t)$ satisfies the following conditions:

$$\int _ { - \infty } ^ { + \infty } t ^ { m } \Psi_u (t) d t = 0 \quad ( m = 0 , 1 , \cdots , p - 1 ) $$

Then $\Psi_u (t)$ is said to have a vanishing moment of order p.

$\Psi_u (t)$ has a vanishing moment of order p, which means that the wavelet function $\Psi_u (t)$ is orthogonal to all polynomials below order p.
Reference\cite {LiOrthogonalModeDecomposition} points out that the low-frequency non-oscillating modes of the signal can be fitted with polynomials. That is, the wavelet function $\Psi_u (t)$ with a p-order vanishing moment helps to ignore the low frequency modes in the original signal and focus only on the high frequency modes. The wavelet function with higher vanishing moment has better localization in frequency domain.

\vspace{1mm}
\textbf{Theorem 2:} $\forall$  finite length real sequence:
$$\boldsymbol{u}=[u(-l),u(-l+1),\hdots,u(0),\hdots,u(l-1),u(l)]$$

Thinking of $\boldsymbol{u}$ as a time series, the sampling period of elements in $\boldsymbol{u}$ is $\Delta$, the total length of time is $T=2l\Delta$. We also assume $\sum_{k=-l}^{l}{u(k)=0}$. Based on the seed sequence $\boldsymbol{u}$, the wavelet function $\Psi_u (t)$ is generated according to formula (\ref{eq:welet_construct1}). Then the necessary and sufficient condition for $\Psi_u (t)$ to have the vanishing moment of order p is

\begin{equation}
\label{eq:welet_construct12}
    \sum_{k=-l}^{l}{k^{m}u(k)}=0\quad ( m = 0 , 1 , \cdots , p - 1 )
\end{equation}

\textbf{Proof:}
It can be seen from equation (\ref{fourier_psi}) that the Fourier transform $F(\omega)$ of the wavelet function $\Psi_u (t)$ is continuous and infinitely differentiable in the open interval $(-\Omega_{\Delta},\Omega_{\Delta})$ in the frequency domain.
From formula (\ref{u_delta})(\ref{u_u}), we can also  know that if
$\omega\in(-\Omega_{\Delta},\Omega_{\Delta})$ then:
\begin{equation}
\label{eq:fouier_simple}
    F(\omega)= \Delta\sum_{k=-l}^{l}{u(k) e^{-j\omega k\Delta}}
\end{equation}

According to the sampling theorem, the bandwidths of the signals we discuss are all located in the interval $(-\Omega_{\Delta},\Omega_{\Delta})$.
So using the frequency domain differential property of the Fourier transform, we can get following:

$$ F(\omega) = \int_{-\infty}^{\infty} \Psi_u(t) e^{-j\omega t} dt $$

\begin{align}  
\label{eq:welet_construct13}
\begin{split}
  \int _ { - \infty } ^ { + \infty } t ^ { m } \Psi_u (t)& e^{-j\omega t}d t = j^m\frac{d^mF(\omega)}{d\omega^m}\\ 
  &( m = 0, \cdots , p - 1 )
\end{split}   
\end{align} 

\vspace{2mm}
Substituting equation (\ref{eq:fouier_simple}) into equation (\ref{eq:welet_construct13}), yields:

\begin{align}  
\label{eq:welet_construct14}
\begin{split}
    \int _ { - \infty } ^ { + \infty } t ^ { m } \Psi_u (t) e^{-j\omega t}d t = & \Delta^{m+1}\sum_{k=-l}^{l}{k^mu(k)} e^{-j\omega k\Delta} \\
    &( m = 0 , 1 , \cdots , p - 1 )
\end{split}    
\end{align}  

In equation (\ref{eq:welet_construct14}), let $\omega\to0$ yield:

\begin{align}  
\label{eq:welet_construct15}
\begin{split}
  \int _ { - \infty } ^ { + \infty } t ^ { m } \Psi_u (t)d t = &\Delta^{m+1}\sum_{k=-l}^{l}{k^mu(k)}\\  
  &( m = 0 , 1 , \cdots , p - 1 )               
\end{split}    
\end{align}  

According to equation (\ref{eq:welet_construct15}), we know that the necessary and sufficient condition for wavelet $\Psi_u (t)$ to have the vanishing moment of order p is:
\begin{equation}
\label{eq:welet_construct16}
    \sum_{k=-l}^{l}{k^{m}u(k)}=0\quad ( m = 0 , 1 , \cdots , p - 1 )
\end{equation}
Q.E.D.
\vspace{5mm}

\textbf{Deduction 1:} $\forall$  finite length real sequence:
$$\boldsymbol{u}=[u(-l),u(-l+1),\hdots,u(0),\hdots,u(l-1),u(l)]$$

With $\boldsymbol{u}$ as the seed sequence, the wavelet function $\Psi_u (t)$ is constructed according to the formula (\ref{eq:welet_construct1}). The necessary and sufficient condition for $\Psi_u (t)$ to have the vanishing moment of order p is :

\begin{equation} \label{Matrix_multiplication_o}
    \begin{bmatrix}
        1  & \hdots & 1 & \hdots &1 \\
        (-l)^1  & \hdots & 0 & \hdots & l^1\\
        \vdots  & \ddots & \vdots & \ddots  & \vdots\\
       (-l)^{p-2}  & \hdots & 0 & \hdots  & l^{p-2}\\
       (-l)^{p-1}   & \hdots & 0 & \hdots  & l^{p-1}\\
    \end{bmatrix}
    \begin{bmatrix}
        u(-l)  \\
        u(-l+1) \\
        \vdots \\
        u(l-1)\\
        u(l)\\
    \end{bmatrix}    
     =
    \begin{bmatrix}
       0 \\
       0 \\
       \vdots\\
       0 \\
       0 \\
    \end{bmatrix}
\end{equation}
 \vspace{3mm}
 
\textbf{Deduction 2:} $\forall$  finite length real sequence:
$$\boldsymbol{u}=[u(-l),u(-l+1),\hdots,u(0),\hdots,u(l-1),u(l)]$$

With $\boldsymbol{u}$ as the seed sequence, the wavelet function $\Psi_u (t)$ is constructed according to the formula (\ref{eq:welet_construct1}). If $\Psi_u (t)$ has the vanishing moment of order p, then $p<2l+1$.

\section{Random wavelet function}
\label{sec:random_wavelet }

For a finite length random number sequence with a mean of zero, it can also be used as a seed sequence to generate a wavelet function according to the formula (\ref{eq:welet_construct1}). We call such a wavelet function a random wavelet function.

\begin{figure}[ht] 
    \centering 
    \subfloat[]{
        \includegraphics[width=6.5cm,height=3.1cm]{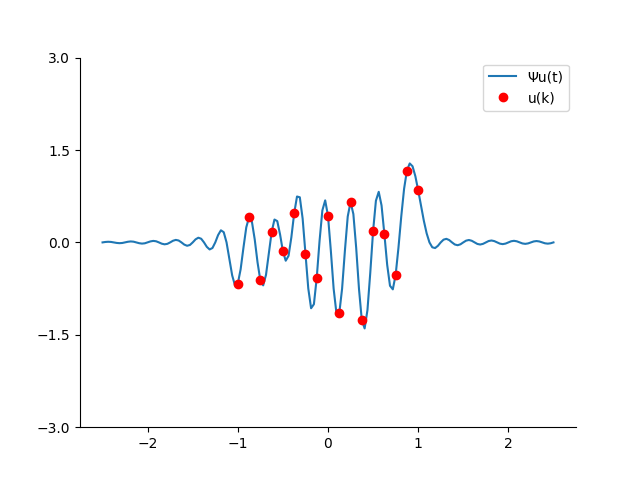}}%
    \hspace{3mm} 
    \subfloat[]{
        \includegraphics[width=6.5cm,height=3.1cm]{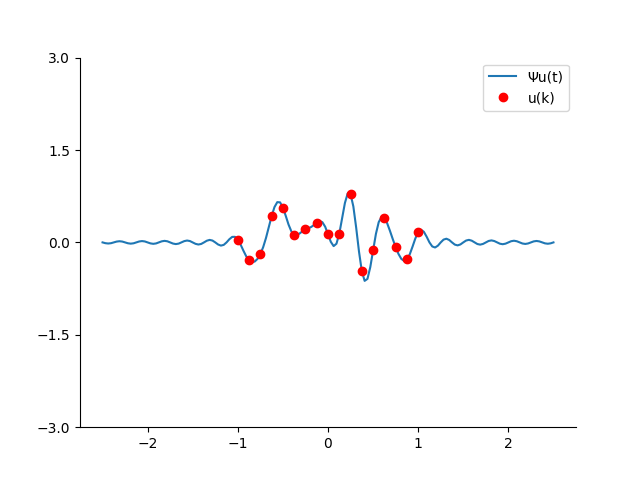}}%
    \hspace*{3mm} 
    \caption{The random seed sequence and the random wavelet}\label{fig:fig_2} 
\end{figure}

The two functions given in example 2 are generated by two random seed sequences with mean 0, both of which meet the conditions of wavelet functions. In Fig.\ref{fig:fig_2}(a), the variance of the seed sequence of the wavelet function is 1. In Fig.\ref{fig:fig_2}(b), the variance of the seed sequence of the wavelet function is 0.6. Our Studies show that the random wavelet function plays an important role in the analysis of stochastic process.

\section{How to make a random wavelet function symmetric?}
\label{sec:random_symmtry }
\vspace{1mm}

Random sequence is generally don't have symmetry, but can be decomposed into odd and even two random subsequence \cite {LiOrthogonalModeDecomposition}.Thus,odd wavelet function and even wavelet function are obtained respectively. Odd wavelet function and even wavelet function have anti-symmetry and symmetry respectively.
\vspace{1mm}

$\forall$ Random seed sequence:
$$ \boldsymbol{u} = [u(-l), u(-l+1),\hdots, u(0),\hdots, u(l-1), u(l)]$$
We arrange $\boldsymbol{u}$ in reverse order, yield: 
$$ \boldsymbol{u_{inv}} = [u(l), u(l-1),\hdots, u(0), \hdots, u(-l+1), u(-l)]$$
The odd subsequence $\boldsymbol{u_o}$ and even subsequence $\boldsymbol{u_e}$ can be obtained according to the following formula:
\begin{align} \label{orth_decomp11}
\begin{split}
    \boldsymbol{u_e} & = 0.5(\boldsymbol{u} + \boldsymbol{u_{inv}}) \\
    \boldsymbol{u_o} & = 0.5(\boldsymbol{u} - \boldsymbol{u_{inv}}) \\
    \boldsymbol{u} & = \boldsymbol{u_e} + \boldsymbol{u_o}
\end{split}
\end{align}
 
\begin{align} 
& \begin{cases}
    u_e(-k) = u_e(k) \\
    u_o(-k) = -u_o(k)
  \end{cases}
& k = 1, 2, ...,   l, 
& \text{     }
& \begin{cases}
  u_e(0) = u(0) \\
  u_o(0) = 0
\end{cases} \nonumber
\end{align}

\begin{figure}[ht]
\centering
\includegraphics[width=6cm,height=3.1cm]{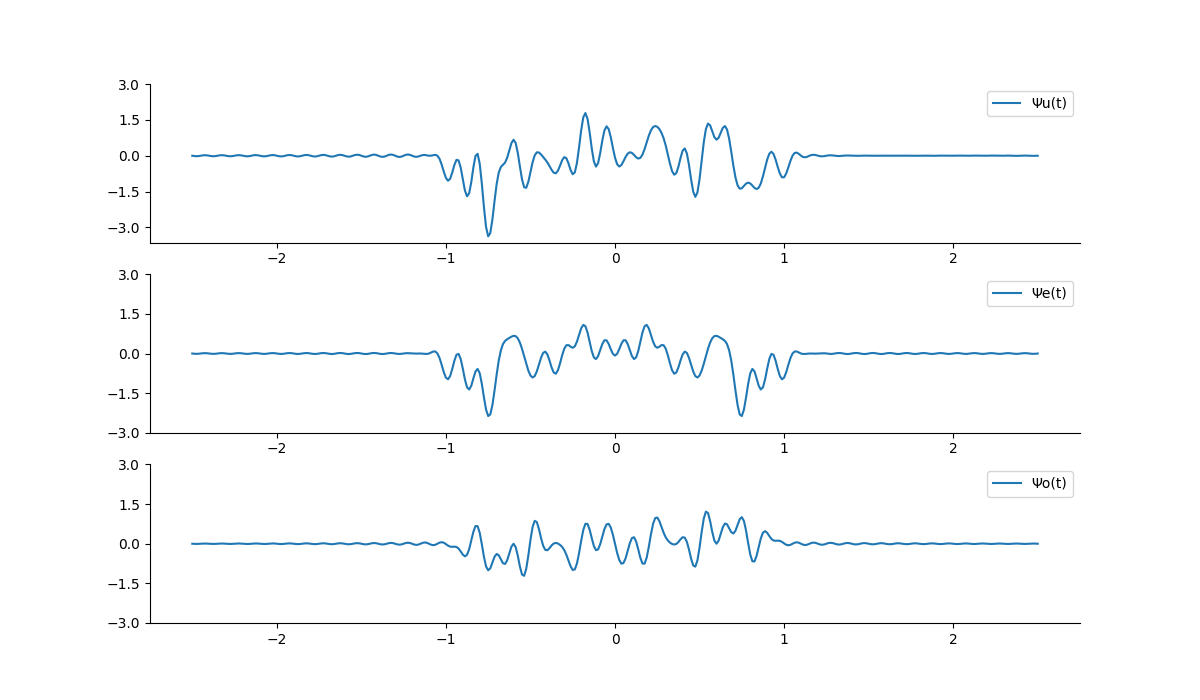}
\caption{The even wavelet and the odd wavelet in example 3}\label{fig:fig_3} 
\end{figure}

\textbf{Example 3:} In example 3, the seed sequence $\boldsymbol{u}$ is a random sequence of length 41, with a mean of 0 and a variance of 1.
The seed sequence $\boldsymbol{u}$ is regarded as a time series signal, its distribution interval is [-1,1], the sampling period is ${\Delta}=0.05s$, and the upper band of the sampling signal is $\Omega_{\Delta}=20\pi$.
The wavelet function generated by a random seed sequence $\boldsymbol{u}$ is a function defined over the field of real numbers.
The seed sequence $\boldsymbol{u}$ is decomposed into the odd subsequence $\boldsymbol{u_o}$ and the even subsequence $\boldsymbol{u_e}$ according to the formula (\ref{orth_decomp11}).
Fig.\ref{fig:fig_3} shows the wavelet function $\Psi_u(t)$ generated from the seed sequence $\boldsymbol{u}$, even wavelet $\Psi_e(t)$ generated from the seed sequence $\boldsymbol{u_e}$, and odd wavelet $\Psi_o(t)$ generated from the seed sequence $\boldsymbol{u_o}$. Even wavelet function has symmetry, odd wavelet function has anti-symmetry.

\vspace{-2mm}

\section{Vanishing moment of random wavelet function}
\label{sec:random_vanish_moment}
A random wavelet function can also have a vanishing moment of a specified order as long as the its seed sequence has a sufficient length. If we want to generate a random wavelet with vanishing moments of order $p$, we need a seed sequence of length $n(n=2l+1,n>p)$.  We can compose the seed sequence by the following steps:

\begin{enumerate}
  \item If the length of the seed sequence is $n$ and a random wavelet function with vanishing moments of order $p$ is required to be constructed, then we need to generate a random sequence of length $l_R(l_R=[\frac{n-p}{2}])$:  
  $$\boldsymbol{u}_R=[u_R(0),u_R(1),\hdots,u(l_R-1)]$$ 
 \item The square matrix $M$ is formed by the $p$ columns in the middle of the coefficient matrix of the linear system (\ref{Matrix_multiplication_o}), that is:
\begin{equation} \label{Matrix_multiplication_o11}
     M=     
       \begin{bmatrix}
        1  & \hdots & 1 & \hdots &1 \\
        (-l+l_R-1)^1  & \hdots & 0 & \hdots & (l-l_R+1)^1\\
       \vdots & \ddots & \vdots & \ddots &\vdots \\
       (-l+l_R-1)^{p-2}  & \hdots & 0 & \hdots & (l-l_R+1)^{p-2} \\
       (-l+l_R-1)^{p-1}  & \hdots & 0 & \hdots &(l-l_R+1)^{p-1}\\
     \end{bmatrix}  \nonumber     
\end{equation}

\item Taking matrix $M$ as coefficient matrix and $C_{const}$as constant term vector, the system of linear equations is established, that is:
\begin{equation}
\label{eq:liner}
  M\boldsymbol{x}^T=C_{const}^T       
\end{equation} 

In the system of linear equations (\ref{eq:liner}), $\boldsymbol{x}$ is an solution vector:
$$\boldsymbol{x}=[x_0,x_1,\hdots,x_{p-1}]$$  $C_{const}=[c_0,c_1,\hdots,c_{p-1}]$ is a constant term vector:
\begin{equation}
  c_i =
    \begin{cases}
      -2\sum_{j=0}^{l_R-1}{(-l+j)^i u_R(j)} & \text{$i = 0, 2, \hdots, p-1$ }\\
      0 & \text{$i = 1,3, \hdots p-2 $}.           \nonumber
    \end{cases}       
\end{equation} 

\item To solve the linear system of equations (\ref{eq:liner}), get the solution vector $\boldsymbol{x} = [x_0 x_1,\hdots,x_{p - 1]}$. It is easy to prove that $\boldsymbol{x}$ is symmetric, that is:
$$x_i=x_{p-i-1} \hspace{0.9cm}    i=0,1,\hdots,[\frac{p-1}{2}]$$
\item The random sequence $\boldsymbol{u_R}$ is rearranged in reverse order to obtain a random sequence in reverse order:
$$\boldsymbol{u}_{RR}=[u_R(l_R-1),u_R(l_R-2),\hdots,u(0)]$$
\item Concatenating $\boldsymbol{u}_R$, $\boldsymbol{x}$, $\boldsymbol{u}_{RR}$ to form a new sequence: 
$$\boldsymbol{u}=\boldsymbol[\boldsymbol{u}_R,\boldsymbol{x},\boldsymbol{u}_{RR}]$$

Based on the sequence $\boldsymbol{u}$, a symmetric random wavelet function $\Psi_u$ can be constructed from formula (\ref{eq:welet_construct1}).
\end{enumerate}
\vspace{3mm}

\begin{figure}[ht] 
    \centering 
    \subfloat[]{
        \includegraphics[width=4.5cm,height=3.1cm]{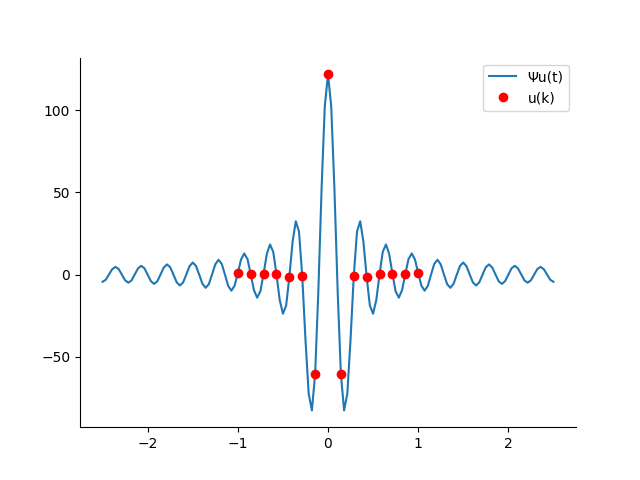}}%
    \hspace*{-3mm} 
    \subfloat[]{
        \includegraphics[width=4.5cm,height=3.1cm]{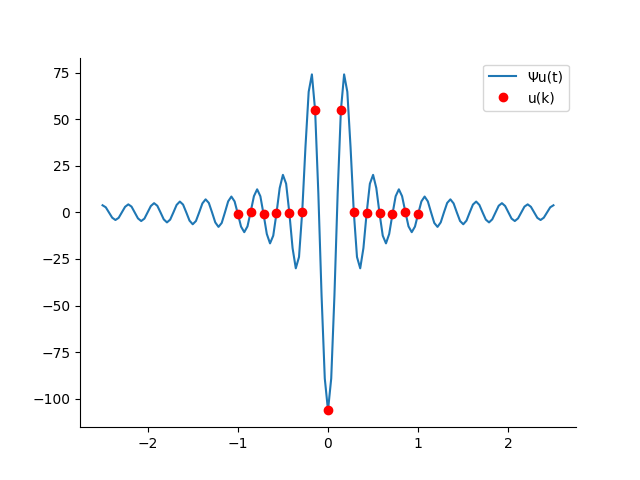}}%
    \hspace*{3mm} 
    \caption{Random seed sequence and random wavelet Function }\label{fig:fig_4} 
\end{figure}

\textbf{Example 4:} In example 4, the length of the seed sequence is 15, and the random wavelet function is required to have a vanishing moment of order 3. The seed sequence $\boldsymbol{u}$ is treated as the sampling signal.Its time domain distribution is [-1,1], sampling period is ${\Delta}=0.05s$, and the upper band of the sampling signal is $\Omega_{\Delta}=20\pi$.
Two cases are given in figures \ref{fig:fig_4}(a) and \ref{fig:fig_4}(b). In both cases, the randomly generated sequence has a mean of 0 and a variance of 1.
The seed sequence $\boldsymbol{u}$ of the wavelet function is marked with red dots in the figure, where the 6 points on the left and right sides are the random sequence $\boldsymbol{u}_R$ and its reverse order $\boldsymbol{u}_{RR}$,
The three dots in the middle is the solution vector $\boldsymbol{x}$ obtained by solving the system of linear equations (\ref{eq:liner}).

There is a significant difference between Fig.\ref{fig:fig_4}(a) and Fig.\ref{fig:fig_4}(b). This is because the randomness of the sequence $\boldsymbol{u}_R$ leads to the randomness of the solution sequence $\boldsymbol{x}$, which leads to a significant difference between the two Figures. But both figures show the wavelet function with  vanishing moment of order 3.

\section{Conclusion and prospect}

The wavelet function refers to the short wave shape of finite length and fast decay, and has an oscillating shape. It is obvious that the interpolation function constructed based on finite length real number sequence conforms to the shape of wavelet. 
We call such finite-length real numbers sequence generating wavelets as seed sequence, and the corresponding wavelet function is called seed sequence wavelet function. The seed sequence wavelet function is continuous and differentiable. It also has limited support sets in both the time and frequency domains.The seed sequence determines the properties of the wavelet function.The zero mean of the seed sequence is equivalent to the "admissibility condition" of the wavelet function. The higher-order vanishing moment of the wavelet function is also related to the corresponding properties of the seed sequence. 

Random wavelet function can be generated based on random sequence with zero mean. As long as the random sequence has sufficient length, the random wavelet function can also meet the corresponding regularity requirements.

Our further research shows that the random wavelet function is of great significance and has the potential to be a tool for the study of stochastic processes.

\end{document}